\documentclass[preprint*]{JHEP3} 

\JHEPspecialurl{http://jhep.sissa.it/JOURNAL/JHEP3.tar.gz}

\usepackage{epsfig,multicol}
\usepackage{amsmath,amssymb}

\newcommand\fverb{\setbox\pippobox=\hbox\bgroup\verb}
\newcommand\fverbdo{\egroup\medskip\noindent%
                        \fbox{\unhbox\pippobox}\ }
\newcommand\fverbit{\egroup\item[\fbox{\unhbox\pippobox}]}
\newbox\pippobox

\title{Another approach to local 
cohomology problem \\in abelian lattice gauge theories
}

\author{
Daisuke Kadoh and Yoshio Kikukawa\\
Nagoya University, Nagoya 464-8602, Japan\\
E-mail: \email{kadoh@eken.phys.nagoya-u.ac.jp}, 
        \email{kikukawa@eken.phys.nagoya-u.ac.jp}}

\abstract{

A new technique is proposed to classify a topological field in abelian 
lattice gauge theories. We perform the classification by regarding the 
topological field as a local composite field of the gauge field tensor
instead of the vector potential associated to an admissible gauge
field. Our method reproduces the result obtained by
the ordinary method in the infinite four-dimensional lattice and 
can be extended to arbitrary higher dimensions.
It also works in the direct cohomological analysis on a finite lattice.
}


\begin{document} 

\pagebreak
\section{\label{sec:intro}Introduction}
The classification of topological fields on the lattice is an 
important and interesting question.
In the ordinary compact formulation of lattice gauge theories,
it is not possible to express the topology of gauge fields 
because any link variables can be deformed to unity continuously.
However  
if one considers the class of lattice gauge fields which satisfy 
the admissibility condition\footnote{
Through out this paper, we set the lattice spacing unity,
$a=1$.},
\begin{eqnarray}
\label{eq:admissibility-1}
&& \parallel 1-
U(x,\mu)U(x+\hat\mu,\nu)U^{-1}(x+\hat\nu,\mu)U^{-1}(x,\nu)\parallel
<  \epsilon,
\end{eqnarray}
where $\epsilon$ is sufficiently small positive number, 
then the nontrivial topology of lattice gauge fields emerges 
and a topological invariant can be defined as a smooth, local,
gauge-invariant function of link variables even when lattice spacing 
is finite.\cite{Luscher:1981zq,
Phillips-Stone,Phillips:1990kj}
The density of the invariant is a topological field 
in the sense that its local variation with respect to the gauge 
field sums up to zero:
\begin{equation}
\label{eq:topological-nature}
\sum_x  \delta q(x) = 0 .
\end{equation}
As a consequence of the topological nature plus
the locality and the gauge-invariance,
such topological fields may be classified by the Chern characters
modulo trivial divergence term
as in the continuum theory.
And this local cohomology problem is closely related to the question
how to achieve the exact cancellation of gauge anomalies in chiral 
lattice gauge
theories.\cite{Alvarez-Gaume:1983cs,
Luscher:1998kn,Luscher:1998du,Luscher:1999un,
Adams:2000yi,
Fujiwara:1999fi,Fujiwara:1999fj,Suzuki:2000ii,Igarashi:2000zi,
Luscher:2000zd,Kikukawa:2000kd,Kikukawa:2001mw,
Igarashi:2002zz,Kadoh:2003ii,Kadoh:2004uu}

In the case of abelian gauge group, it has been shown 
by L\"uscher\cite{Luscher:1998kn} that 
any topological field on the four-dimensional infinite lattice
is classified uniquely in the following form:
\begin{eqnarray}
q(x)&=& \alpha + \beta_{\mu\nu} F_{\mu\nu}(x)+
\gamma \, 
\epsilon_{\mu\nu\rho\sigma}F_{\mu\nu}(x)F_{\rho\sigma}(x+\hat\mu+\hat\nu)
+\partial^*_{\mu}k_{\mu}(x),
\label{eq:Luscher-classification}
\end{eqnarray}
where $F_{\mu\nu}(x)$ is the field tensor, 
$\alpha, \beta_{\mu\nu}$ and $\gamma$ are constants independent of
the gauge fields 
and $k_{\mu}(x)$ is a gauge invariant local current.\footnote{
In the formualtion of abelian chiral gauge theories
using the Ginsparg-Wilson fermions\cite{Ginsparg:1981bj,
Neuberger:1997fp,Neuberger:1998wv,
Hasenfratz:1998ri,Hasenfratz:1998jp,
Hernandez:1998et}, 
the gauge anomaly is 
identical to the chiral anomaly, which is given by 
Jacobian of the exact chiral transformation.\cite{Luscher:1998pq,
Kikukawa:1998pd, Fujikawa:1998if, Adams:1998eg, Suzuki:1998yz, Chiu:1998xf} 
It satisfies the index theorem and its local density is a topological 
field. Then using the result eq.~(\ref{eq:Luscher-classification}), 
one can establish the exact cancellation 
of the gauge anomaly.\cite{Luscher:1998du} 
} 
This result can be extended to arbitrary higher 
dimensions.\cite{Fujiwara:1999fj} 
In the proof of the result, in the ordinary approach, 
a vector potential is introduced for each admissible gauge field so that 
it represents the original link variables and the field tensor as 
\begin{eqnarray}
&&  {\rm e}^{i A_\mu(x)} = U(x,\mu), \qquad  |A_\mu(x)| \le
\pi(1+8\|x\|),  \\
&& \partial_\mu A_\nu(x) - \partial_\nu A_\mu(x) = F_{\mu\nu}(x) ,
\end{eqnarray}
and the topological field is regarded as a gauge-invariant 
function of the vector potential.

The aim of this paper is to point out that
the introduction of the vector potential is not actually necessary
to prove the result eq.~(\ref{eq:Luscher-classification}).
We will show that it is indeed possible to give a proof
by regarding the topological field as a gauge-invariant, local, function
of linear independent variables of the field tensor. 
Our new technique can be easily extended to higher dimensional cases. 
It also works in the direct cohomological analysis on the finite-volume
lattice, which is formulated for practical 
uses.\cite{Kadoh:2003ii,Kadoh:2004uu}

The new method based on the field tensor may have some advantages
over the ordinary method using the vector potential. 
In fact, we can simplify the cohomological analysis by
reducing several steps of it, as we will see below.
Moreover, for non-abelian theories, 
a vector-potential-representation 
of the link variables and the field tensor can be rather complex 
because of the non-linear nature of the field tensor. 
If we can do without introducing the vector potential,
the cohomological analysis would be simpler.

This paper is organized as follows. 
In section 2, we recall the definition of 
admissible abelian lattice gauge fields 
and the notion of local composite fiels of gauge fields.
Section 3 is devoted to the description of the basic idea of 
our new method based on the field tensor.
The complete analysis of the topological fields
is given in the next sections 4 and 5.
We conclude in section 6.

\section{Admissible gauge fields and locality}
In this section we give a definition of admissible abelian 
gauge fields on the infinite lattice of dimension $n$. We 
then recall the notion of local composite field of the 
admissible gauge fields.

\subsection{Admissible abelian gauge fields}
\label{subsect:gauge-field}
We consider the compact formulation of abelian lattice gauge theory.
Gauge fields are represented by the link variables 
which take values in the gauge group $U(1)$,
\begin{eqnarray}
&& U(x,\mu) \in {\rm U(1)}, \quad x \in
\mathbb{Z}^n,\quad \mu=1,...,n,
\end{eqnarray}
and gauge transformations are defined to vary the link variables as
\begin{equation}
U(x,\mu) \rightarrow \Lambda(x) U(x,\mu)
 \Lambda(x+\hat\mu)^{-1},\quad \Lambda(x)\in  {\rm U(1)},
\end{equation}
where $\Lambda(x)$ is a gauge transformation function.
The field tensor may be defined through the plaquette variables as
\begin{eqnarray}
&&  F_{\mu\nu}(x) = -i{\rm ln} \, P_{\mu\nu}(x), \quad
\quad - \pi < F_{\mu\nu}(x) \le \pi , \\
&& P_{\mu\nu}(x)=U(x,\mu)U(x+\hat\mu,\nu)U(x+\hat\nu,\mu)^{-1}
U(x,\nu)^{-1}.
\end{eqnarray}

We then impose the so-called admissibility condition given by\footnote{
In two-dimensions, we adopt the following admissibility condition;
$\vert F_{\mu\nu}(x) \vert  < \pi $
 }
\begin{equation}
\vert F_{\mu\nu}(x) \vert  < \frac{\pi}{3} \ \ \ \ \ \ {\rm for \ all} \ \ 
x,\mu,\nu. 
\label{eq.admissibility-2} 
\end{equation}
In general, the field tensor satisfies a lattice counterpart of the 
Bianchi identity,
\begin{equation}
{\rm e}^{i(\,
 \partial_{\mu}F_{\nu\rho}(x)+\partial_{\nu}F_{\rho\mu}(x)
+\partial_{\rho}F_{\mu\nu}(x)\,)}=1
 \qquad  {\rm for  \ } \quad \mu,\nu,\rho=1,...,n,
\end{equation}
where $\partial_{\mu}$ denotes the forward nearest-neighbor 
difference operator.
This implies that the exterior difference of the field tensor
is equal to an integer multiple of $2\pi$. 
However, for admissible gauge fields, it vanishes identically,
\begin{equation}
 \partial_{\mu}F_{\nu\rho}(x)+\partial_{\nu}F_{\rho\mu}(x)
+\partial_{\rho}F_{\mu\nu}(x) =0
 \qquad  {\rm for} \quad \mu,\nu,\rho=1,...,n.
\label{eq:Bianchi-identity}
\end{equation}  
Namely, the usual Bianchi identity holds exactly.

\subsection{Local composite field of admissible gauge fields}
Under the admissibility condition, the 
link variables are not entirely independent. 
Then the locality with respect to the gauge field is not quite trivial.
Here we adopt the definition of local composite fields of 
the admissible gauge fields, which is given in \cite{Luscher:1998kn}.

\vspace{1em}
\noindent
{\it Local composite fields in the infinite lattice}:\\
we refer a composite field $\phi(x)$ on the infinite lattice 
as local 
if $\phi(x)$ is given by the series expansion, 
\begin{eqnarray}
\phi(x)=\sum_{k=1}^{\infty}\phi_k(x) ,
\label{eq:series-expansion}
\end{eqnarray}
where $\phi_k(x)$ are strictly local fields with a localization
range proportional to $k$ and the derivatives
$\phi_k(x;y_1,\nu_1;\cdots;y_m,\nu_m)$ with respect to the gauge 
field variables $U(y_1,\nu_1)$, $\cdots$, $U(y_m,\nu_m)$ exist 
and are bounded by
\begin{eqnarray}
|\phi_k(x;y_1,\nu_1;\cdots;y_m,\nu_m)|\le c_m k^{p_m}e^{-\theta k},
\label{eq:exponentially-1}
\end{eqnarray}
where the constants $c_m,p_m >0$ and $\theta >0$ are independent of the
gauge field. 

\vspace{1em}
\noindent
From this definition, it is obvious that the infinite sum
eq.~(\ref{eq:series-expansion}) is well-defined and 
the derivatives of $\phi(x)$
satisfy the following locality property,
\begin{eqnarray}
|\phi(x;y_1,\nu_1;\cdots;y_m,\nu_m)| \le d_m
 (1+\|x-z\|^{p_m})\,e^{-\theta \|x-z\|} ,
\label{eq:locality property}
\end{eqnarray} 
where $z$ is chosen from $y_1,\cdots,y_m$ so that 
$\|x-z\|$ is the maximum.
The constant $d_m >0$ is independent of the gauge field.

\section{New approach to local cohomology problem}

In this section, 
we introduce a new technique in which 
the topological field 
on the lattice is regarded as a local composite field in terms of 
the linear independent components of the field tensor. 
We first explain how to choose the linear independent components of the field 
tensor. We also argue that the locality property of the topological field
is maintained through this change of variables. Then we describe
the basic idea of the cohomological analysis based directly on the 
field tensor. The complete analysis of the topological field
is given in the next sections.

\subsection{Field tensor as independent variables}

If a composite field of link variables is gauge-invariant, 
it can be regarded as a function of the field tensor
instead of the link variables.
But, the components of the field tensor are 
not linear independent due to the Bianchi identity 
eq.~(\ref{eq:Bianchi-identity}), and 
in order to treat the field tensor as the independent variables,
we must specify a set of the linear independent components
of the field tensor.
A simple way to do this is to use an appropriate complete gauge fixing.
Because, in the course of solving the gauge-fixing condition,
a certain set of the independent components of the field tensor 
is automatically selected and the gauge-fixed link variables are 
represented as a product of these independent plaquette variables. 

For our purpose, we adopt the complete axial gauge with 
the reference point set to the origin. In this gauge, 
the link variables are fixed as follows:
\begin{eqnarray}
U^g(x,\mu)|_{x_1=\cdots=x_{\mu-1}=0}=1 \quad \quad
{\rm for}\,\,\,\mu=1, \ldots ,n.
\label{eq:hat U constraint}
\end{eqnarray}
The gauge transformation function of this complete axial gauge is given by 
the product of the link variables on the path $(0,x)_{\rm p}$
which starts from the origin and ends at $x$ 
along which the coordinate $x_i$ 
increases (or decreases) in the descending order $i=n,n-1,\ldots,1$: 
\begin{equation}
\Lambda(x)=\prod_{U\, 
{\in}\, 
(0,x)_{\rm p}} \,U(z,\mu).
\label{eq:gauge fix}
\end{equation}
Non-trivial components of the gauge-fixed link variables are obtained by 
solving  
\begin{equation}
U^g(x+\hat\mu,\nu) U^g(x,\nu)^{-1} = {\rm exp}\{i F_{\mu\nu}(x)\} \ \ \
 \ \ \ 
{\rm for}\,\,\,\nu > \mu
\end{equation}
in the order $\mu=n-1,\ldots,1$.
Then these variables are expressed 
by the product of the plaquette variables which are specified 
by the way to solve the gauge-fixing condition. Explicitly they 
are given as 
\begin{eqnarray}
U^g(x,\mu)= \prod_{ F \,\in\, S(0,x;\mu)} \,{\rm exp}({i 
F_{\rho\sigma}(z)}) \ \ \ \ \ \ {\rm for \,\, all} \ \ x,\mu, 
\label{eq:field-tensor-rep}
\end{eqnarray}
where $S(x_0,x;\mu)$ is a set of variables of the field tensor 
which is contained in the minimal surface with the 
boundary $(x_0,x)_{{\rm p}}+(x,x+\hat\mu)_{{\rm p}}-
(x+\hat\mu,x_0)_{{\rm p}}$. 
The components of the field tensor which appears in the above formula 
must be independent degrees of freedom of the gauge field. 
The other components of the field tensor are to be computed 
from the gauge-fixed link variables, or using the Bianchi identity.
Note also that this choice of the independent components of the
field tensor depends on the reference point of the complete axial gauge. 
Thus we can specify the set of the independent components of the field 
tensor with a reference point $x_0$ as
\begin{eqnarray}
\mathfrak{F}_{{}_{x_0}}
& \equiv & \{\,F_{\rho\sigma}(z)\,|\, F_{\rho\sigma}(z)\in
 S(x_0,x,\mu),x\in \mathbb{Z}^n,\mu=1,\ldots,n\}\\
&=& 
\{\,F_{\rho\sigma}(z)|_{z_1={x_0}_1,\cdots,z_{\mu-1}={x_0}_{\mu-1},},\, 
\rho,\sigma=1,\ldots,n,\, \sigma>\rho,\} .
\end{eqnarray}

\subsection{Locality property in terms of the field tensor}
\label{subsect:lolcality}

From the discussion in the previous subsection, 
we can see that any gauge-invariant composite field $\varphi(x)$ of 
link variables can be regarded 
as a composite field of the independent components of the field tensor,
\begin{equation}
 \varphi(x)[{U}]
=\varphi(x)[{U^{g}}]
=\varphi(x)[{F} \in \mathfrak{F}_{{}_{x_0}}] .
\end{equation}
We next argue that in this change of variables 
the locality property of the composite field can be maintained
by the suitable choice of the  reference point.

Let $\phi(x)$ be a gauge-invariant local composite field 
of the link variables. 
Then we can see that 
the derivatives with respect to the gauge-fixed link variables 
satisfy eq.~(\ref{eq:locality property}).
Next we consider the derivatives of $\phi(x)$ 
with respect to the independent components of the field tensor with
the reference point $x$, that is, $\mathfrak{F}_{{}_x}$. 
According to eq.~(\ref{eq:field-tensor-rep}), 
the variation of the component of the field tensor at some point $y$
generates many variations of the gauge-fixed link variables. But 
these variations occur at the points which are further than $y$ 
from $x$ in the taxi driver distance. Then, by using the chain rule, 
we can show that 
the derivatives $\phi(x;y_1,\mu_1,\nu_1;\cdots;y_m,\mu_m,\nu_m)$ 
with respect to the field tensor 
$F_{\mu_1\nu_1}(y_1),\cdots,F_{\mu_m\nu_m}(y_m) \in
\mathfrak{F}_{{}_{x}}$ satisfy the bound
\begin{eqnarray}
|\phi(x;y_1,\mu_1,\nu_1;\cdots;y_m,\mu_m,\nu_m)|
\le d'_m (1+\|x-z\|^{q'_m})\,e^{-\theta \|x-z\|} ,
\label{eq:locality property_F}
\end{eqnarray} 
where $z$ is chosen from $y_1,\cdots,y_n$ so that $\| x-z \|$ is
maximal and 
the constants $d'_m,q'_m >0$ independent of the gauge field. 

Conversely, let us assume that $\phi(x)$ has the locality
property 
with respect to the independent components of the 
field tensor, eq.~(\ref{eq:locality property_F}). 
Since the field tensor is a local field of the link variables, 
it immediately follows that the composite field 
has the locality property 
with respect to the link variables, eq.~(\ref{eq:locality property}).
Thus the locality property of a gauge-invariant local composite field 
does not depend on the choice of the sets of the independent variables.

\subsection{Field tensor-based cohomological analysis}

We now describe how to perform the cohomological analysis of 
topological fields with the field tensor.
Let us consider a topological field $q(x)$, which is 
a gauge-invariant local composite field of admissible gauge
fields and satisfy the property,
\begin{equation}
\label{eq:topological-nature-2}
\sum_{x \in {\mathbb{Z}}^n}  \delta q(x) = 0 ,
\end{equation}
under the local variation of the link variables.
We regard the value of $q(x)$ at the point $x$ 
as the local function of the 
linear independent components of the field tensor in $\mathfrak{F}_{{}_{x}}$. 
Scaling those components by the
parameter $t \in [0,1]$ and differentiating and integrating $q(x)$ 
with respect to the parameter $t$, we obtain 
\begin{eqnarray}
\label{eq:q-theta-F}
q(x)=\alpha + \frac{1}{2} \displaystyle \sum_{y\in
 \mathbb{Z}^n}\theta_{\mu\nu}(x,y) F_{\mu\nu}(y),
\ \ \ \ \ \ \theta_{\mu\nu}(x,y)= \left.\int_0^1 {\rm d}t\, 
\hat\theta_{\mu\nu}(x,y)\right|_{ F \rightarrow t F} ,
\end{eqnarray}
where $\alpha$ is a constant that is independent of the gauge field and
$\hat \theta_{\mu\nu}(x,y)$ is given by
\begin{equation}
\hat\theta_{\mu\nu}(x,y)
=\left\{ 
\begin{array}{cl}
\displaystyle 
\left. \frac{\partial q(x)}{\partial F_{\mu\nu}(y)}\right.
&\qquad {\rm for} \,\, F_{\mu\nu}(y) \in \mathfrak{F}_{{}_x},\\
0 
&\qquad {\rm for}\,\, F_{\mu\nu}(y)\notin \mathfrak{F}_{{}_x}.\\
\end{array}
\right. 
\end{equation}
Note that the dependent components of the field tensor appears in 
eq.~(\ref{eq:q-theta-F}). These terms do not 
actually contribute to the topological field because 
the coefficients $\hat \theta_{\mu\nu}(x,y)$ of the dependent components
vanish identically. But for later convenience we include them as 
dummy variables and the summation over the indices $y,\mu,\nu$
are taken for all possible values.

In order to derive the constraint on 
the bi-local field $\theta_{\mu\nu}(x,y)$, which
follows from the topological property of $q(x)$, 
we consider an infinitesimal variation of the link variables,
\begin{equation}
\delta_\eta U(x,\mu) = i \eta_\mu(x) U(x,\mu).
\label{eq:variation}
\end{equation}
This causes the variations of both the independent and
dependent components of the field tensor as 
\begin{equation}
\delta_\eta F_{\mu\nu}(y) = \partial_\mu \eta_\nu(y) - \partial_\nu
 \eta_\mu(y) .
\label{eq:variation_F}
\end{equation}
Then the topological field $q(x)$ changes as 
\begin{eqnarray}
\delta_\eta q(x)
= \displaystyle \sum_{y\in
 \mathbb{Z}^n}\hat\theta_{\mu\nu}(x,y) \delta_\eta F_{\mu\nu}(y). 
\end{eqnarray}
By using eq.(\ref{eq:variation_F}) and the integration over $t$,
the topological property of $q(x)$ leads to the following constraint, 
\begin{eqnarray}
\displaystyle \sum_{x \in \mathbb{Z}^n} 
\theta_{\mu\nu}(x,y)\overleftarrow{\partial^*_{\nu}}= 0 , 
\end{eqnarray}
where $\overleftarrow{\partial^*_{\nu}}$ is the backward
nearest-neighbor difference operator with respect to $y$.

We note that the above relation is exactly same as  that obtained
in the course of the original cohomological analysis.\cite{Luscher:1998kn} 
Once we obtain this relation, we can immediately see that 
the same argument as in the original analysis
using the Poincar\'e lemma applies and it leads to the same first-step 
result, lemma 6.1. in \cite{Luscher:1998kn}.  The second-step result, 
lemma 6.2 in \cite{Luscher:1998kn}, can be also obtained by this method. 
Thus we can see that our method reproduces the result 
of the original cohomological analysis in \cite{Luscher:1998kn} and 
that the cohomological classification can be achieved
without introducing the vector potential representation of 
admissible gauge fields. 

In the following sections, we perform systematic cohomological analysis 
based on the direct use of the field tensor 
for topological fields in arbitrary n-dimensional infinite lattices
and finite-volume lattices. 

\section{Field tensor-based analysis in 
$n$-dimensional infinite lattice}
\label{sec:cohomological-analysis-infiniteV}

\subsection{A lemma for the difference operators}
\label{subsec:lemma-infiniteV}

In the following cohomological analysis, we require 
a version of the Poincar\'e lemma on difference operators.
For this purpose, we first introduce the differential 
forms defined on the n-dimensional lattice. 
A general $k$-form is defined through 
\begin{equation}
f(x) = \frac{1}{k!} f_{\mu_1\cdots\mu_k}(x) 
{\rm d}x_{\mu_1}\cdots {\rm d}x_{\mu_k},
\end{equation}
where $f_{\mu_1\cdots\mu_k}(x)$ is totally anti-symmetric
field and  ${\rm d}x_{\mu_1}, \cdots, {\rm d}x_{\mu_k}$ satisfy
the Grassmann algebra. We denote a
linear space of $k$-forms by $\Omega_k$.
For $k < 0$ or $k > n$, $\Omega_k=\emptyset$.

An exterior difference operator ${\rm d}:\Omega_k 
\rightarrow \Omega_{k+1}$ and a divergence operator ${\rm d^\ast}:\Omega_k 
\rightarrow \Omega_{k-1}$ may be defined by
\begin{eqnarray}
&&{\rm d} f(x) = \frac{1}{k!} \partial_\mu 
f_{\mu_1\cdots\mu_k}(x) 
{\rm d}x_\mu {\rm d}x_{\mu_1}\cdots {\rm d}x_{\mu_k},\\
&&{\rm d^\ast} f(x) = \frac{1}{(k-1)!} \partial_\mu^\ast
f_{\mu\mu_2\cdots\mu_k}(x) {\rm d}x_{\mu_2}\cdots {\rm d}x_{\mu_k},
\end{eqnarray}
where $\partial_\mu$ and $\partial_\mu^\ast$ denote the forward and 
backward nearest-neighbor difference operators, respectively. 
We then introduce more general class of difference operators,  $L:\Omega_l
\rightarrow \Omega_{k}$,  which are defined by 
\begin{eqnarray}
\label{eq:L}
Lf(x)=\frac{1}{k!l!}{\rm d}x_{\mu_1}\cdots {\rm
 d}x_{\mu_k}  \sum_{y\in\mathbb{Z}^n}
L_{\mu_1\cdots\mu_k\,,\,\nu_1\cdots\nu_l}(x,y)f_{\nu_1\cdots\nu_l}(y).
\end{eqnarray}
We assume that $L_{\mu_1\cdots\mu_k\,,\,\nu_1\cdots\nu_l}(x,y)$ 
has a compact support including $x$ with respect to $y$,  
and vice versa. We also define simple
multiplication operations by using the following abbreviations of the
summation of the lattice index:
$\sum_1 L:\Omega_l
\rightarrow \Omega_{k}$ stands for the total summation
of the kernel function $L_{\mu_1\cdots,\nu_1 \cdots}(x,y)$
over $x$,
\begin{eqnarray}
\label{eq:sum-L}
\sum\underset{^1}{} Lf(x)=\frac{1}{k!l!}{\rm d}x_{\mu_1}\cdots {\rm
 d}x_{\mu_k}  \sum_{z\in\mathbb{Z}^n}
L_{\mu_1\cdots\mu_k\,,\,\nu_1\cdots\nu_l}(z,x)f_{\nu_1\cdots\nu_l}(x).
\end{eqnarray}

It is also convenient to 
introduce the $\circ$ product 
$:\Omega_p \otimes\Omega_k \rightarrow \Omega_{p-k}$ 
defined by 
\begin{eqnarray}
&&\hspace{-4mm}
\theta \circ f(x) = \frac{1}{(p-k)!k!}{\rm d}x_{\mu_1}\cdots {\rm
 d}x_{\mu_{p-k}}\times\nonumber\\
&&\qquad \qquad \theta_{\mu_1\cdots\mu_{p-k}\nu_{1}\cdots\nu_k}
(\bar x-\hat\nu_1\cdots-\hat\nu_k)
f_{\nu_{1}\cdots\nu_k}(\bar x-\hat\nu_1\cdots-\hat\nu_k),
\end{eqnarray}
where $\bar x = x + \hat 1 + \cdots + \hat n$,
first introduced by Suzuki.\cite{Suzuki-lecture}\footnote{
$x$ is shifted to $\bar x$ so that the expression of the 
Chern characters in the final result assumes the form given 
in eq.~(\ref{eq:final-result-infiniteV}).
}
This product have the crucial property,
\begin{equation}
\label{eq:circle-prod-property}
{\rm d^\ast} \theta \circ = ({\rm d^\ast} \theta) \circ +(-1)^{p-k-1}
 \theta \circ {\rm d},
\end{equation}
when acting on $k$-forms.

In the cohomological analysis of the topological field,
we often encounter a difference operator $L$ 
which satisfies ${\rm d^\ast} L{\rm d}=0$.
The question how 
to express such a difference operator
has been solved originally by Fujiwara et al.\cite{Fujiwara:1999fj}
Here we state the result as a lemma in terms of 
difference operators using the $\circ$ product. 
The proof is given in appendix~\ref{app:lemma-infiniteV}.

\vspace{1em}
\noindent{\bf Lemma~\ref{subsec:lemma-infiniteV}}\\
Let $L:\Omega_l \rightarrow \Omega_k \, (l\not = 0)$ be a difference operator which
satisfies\\
\begin{equation}
\hspace{-2cm}{\rm d^\ast}L{\rm d}=0
\quad {\text{and}} \quad \sum\hspace{-0.5mm}{\underset {1}{}} L {\rm d}=0
\,\,\,\,{\rm \,if\,\,\,}k=0
\end{equation}
Then there exist a form $\theta \in \Omega_{k+l}$ that satisfies ${\rm
d^\ast}\theta=0$ and two difference
operators, $Q_1:\Omega_{l+1} \rightarrow \Omega_k$ and $Q_2:\Omega_l \rightarrow 
\Omega_{k+1}$ such that
\begin{eqnarray}
L=\theta \circ +Q_1 {\rm d}+ {\rm d^\ast} Q_2.
\end{eqnarray}

If the coefficient of $L$ is local with respect to $x$ or $y$ in 
the sense defined in subsect.\ref{subsect:lolcality}, then the
coefficients of $Q_1$ and $Q_2$ are also local with respect to $x$ or
$y$ and $\theta$ is a strict local field. Moreover, 
if $L$ satisfies the following locality
condition instead of the compact support,
\begin{eqnarray}
|L_{\mu_1\cdots\mu_k\,,\,\nu_1\cdots\nu_l}(x,y)|< c(1+\|x-y\|^p){\rm e}^{-\|x-y\|/\varrho},
\end{eqnarray}
where $c,p$ and $\varrho$ are constants independent of the gauge
fields, then $Q_1$ and $Q_2$ satisfy
\begin{eqnarray}
&& | Q_{i\,\,\mu_1\cdots\mu_k\,,\,\nu_1\cdots\nu_l}(x,y)|< c_i(1+\|x-y\|^p){\rm
 e}^{-\|x-y\|/\varrho} \quad \ \ {\rm for}\ \ i=1,2 ,
\end{eqnarray}
where $c_i$ are constants independent of the gauge fields, and $\theta$
is a local field in the sense defined in subsect.\ref{subsect:lolcality}.  

\subsection{Field tensor-based analysis of topological fields 
on the infinite lattice}
\label{subsec:F-based-cohomological-analysis-infiniteV}

We first establish the following lemma through the field-tensor based method. 

\vspace{1em}
\noindent{\bf Lemma~\ref{subsec:F-based-cohomological-analysis-infiniteV}}

A gauge invariant local field $\phi(x) \in \Omega_p$ which satisfies
\begin{eqnarray}
{\rm d^\ast} \phi(x)=0 \,{\rm \,\,and\,\,}\,\sum_{x \in \mathbb{Z}^n}
 \delta\phi(x)=0
\,\,\,\,{\rm \,if\,\,\,}p=0,
\end{eqnarray}
where $\delta$ is any local variation of the admissible gauge fields.
Then there exist two gauge invariant local fields, $\theta(x) \in
\Omega_{p+2}$ which satisfies ${\rm d^\ast}\theta=0$ 
and $\omega(x) \in \Omega_{p+1}$, such that 
\begin{eqnarray}
\phi(x)=c+\theta \circ F(x) + {\rm d^\ast} \omega(x),
\end{eqnarray}
where $c$ is a constant which is independent of the gauge field.\\

\noindent {\sl Proof:}\ \ The gauge invariant local field $\phi(x)$ 
is regarded as a function of the independent components of the field tensors,
$F_{\mu\nu}(y) \in \mathfrak{F}_{{}_x}$. 
We now rescale the independent variables by 
a continuous parameter $t \in [0,1]$, differentiate
$\phi(x)$ by $t$ and integrate it over the region $[0,1]$. Then we obtain
\begin{eqnarray}
\label{eq:difference-operator-L}
\phi(x)=c+ L F(x),\ \ \ \ 
L = \left.\int_0^1 {\rm d}t\, \hat L\right|_{F \rightarrow t F} ,
\end{eqnarray}
where $c=\phi(x)|_{F=0}$ is a constant and a gauge invariant 
difference operator 
$\hat L:\Omega_2 \rightarrow \Omega_p$ is defined by
\begin{eqnarray}
\label{eq:difference-operator-hat-L}
\hat L_{\mu\nu}(x,y)
=\left\{ 
\begin{array}{cl}
\displaystyle\frac{\partial \phi(x)}{\partial F_{\mu\nu}(y)}
\qquad & {\rm for} \,\, F_{\mu\nu}(y) \in
  {\mathfrak {F}}_x,\\
0 \qquad & {\rm otherwise},\\
\end{array}
 \right. 
\end{eqnarray}
The field tensor transforms as
\begin{equation}
\delta_{\eta} F=d \eta
\end{equation}  
under any local variation of the admissible 
gauge fields, eq.(\ref{eq:variation}).
By the assumption of the lemma, $L$ has the following property,
\begin{eqnarray}
{\rm d^\ast}L{\rm d}=0\,{\rm \,\,and\,\,}\,\sum\hspace{-0.5mm}{\underset {1}{}}L {\rm d}=0
\,\,\,\,{\rm \,if\,\,\,}p=0.
\label{eq:premise}
\end{eqnarray} 
These are the premises of the lemma~\ref{subsec:lemma-infiniteV}(b) and 
therefore we have
\begin{equation}
L=\theta \circ +Q_1{\rm d}+{\rm d^\ast}Q_2,
\label{eq:lemma-4.1}
\end{equation}
where $Q_1:\Omega_{3}\rightarrow\Omega_{p}$ and 
$Q_2:\Omega_{2}\rightarrow\Omega_{p+1}$ are gauge invariant difference
operators and $\theta \in
\Omega_{p+2}$ is a gauge invariant field.
Using eq.(\ref{eq:lemma-4.1}) and the Bianchi identity, we obtain 
\begin{equation}
\phi(x)=c+\theta \circ F(x) + {\rm d^\ast} \omega(x),
\end{equation}
where $\omega=Q_2F$ and ${\rm d^\ast}\theta=0$.\quad  $\square$

Now let us consider a gauge invariant, local and topological
field $q(x)$. We apply the above lemma to $q(x)$ by regarding it as 
$\phi(x) \in \Omega_0$.
In this step, we obtain a gauge-invariant, local field 
$\theta \in \Omega_2$ which 
satisfies $d^\ast \theta =0$. This condition for $\theta$ is
just same as the premise of the lemma. 
Next we apply the lemma to $\theta$. 
Repeating this and using eq.~(\ref{eq:circle-prod-property})
and the Bianchi identity, 
we finally obtain the desired result. 
\begin{eqnarray}
\label{eq:final-result-infiniteV}
q(x)= 
\sum_{k=0}^{[n/2]} 
(\cdots((c^{(k)} \circ F)\circ F)\cdots \circ F)
\circ F(x)+{\rm d^\ast}\omega(x),
\end{eqnarray}
where 
in the $k$-th term of the summation
the number of $\circ$ product counts $k$.
Explicitly, 
\begin{eqnarray}
\label{eq:final-resul}
q(x)=
\sum_{k=0}^{[n/2]} \left(\frac{1}{2}\right)^k
 c^{(k)}_{\mu_1,...,\mu_{2k}}\,F_{\mu_1\mu_2}(x)F_{\mu_3
 \mu_4}(x+\hat\mu_1+\hat\mu_2)\times \cdots \ \ \ \  \nonumber \\
\times
 F_{\mu_{2k-1}\mu_{2k}}(x+\hat\mu_1+\cdots \hat\mu_{2k-2})
+{\rm d^\ast}\omega(x) .
\end{eqnarray}

\section{Field tensor-based analysis in 
$n$-dimensional finite lattice}
\label{sec:cohomological-analysis-finiteV}

\subsection{Admissible gauge fields on the finite lattice} 

We define the admissible abelian gauge fields on the finite
$n$-dimensional lattice by imposing 
the periodic boundary condition,
\begin{equation}
U(x,\mu)=U(x+L\hat\nu,\mu)\ \ \ \ \ \ 
{\rm for\,\,}\ \ x\in \mathbb{Z}^n,\,\,\mu,\nu=1,\ldots,n.
\end{equation}
Independent degrees of freedom of the gauge fields are restricted in
the finite box,
\begin{equation}
\displaystyle \Gamma_n=\left\{\,x\, \left|\,-\frac{L}{2}
\le x_{\mu}<\frac{L}{2}-1,\quad {\rm for}\ \ \mu=1,...,n \right.\right
\} ,
\end{equation}
where $L$ is a lattice size.
The space of such gauge fields is devided
into the connected subspaces labeled by the magnetic flux numbers,
\begin{equation}
m_{\mu\nu}=\frac{1}{2\pi}\sum_{s,t=0}^{L-1}\,F_{\mu\nu}(x+\hat\mu+\hat\nu).
\end{equation}
Each subspace is isomorphic to a space of 
the product of U(1) times a contractible space.
\begin{eqnarray*}
\mathfrak{U}[m] \cong U(1)^N \times [\text{a contractible space}],
\end{eqnarray*}
where $N=d+L^d-1$, the number of non-contractible Wilson loops plus
the number of gauge transformations which are independent each
other. In fact, 
the link variables of the gauge field in $\mathfrak{U}[m]$ are
uniquely represented as
\begin{eqnarray}
 U_{[m]}(x,\mu)
=\Lambda(x)U_{[w]}(x,\mu)\Lambda^{-1}(x+\hat\mu)V_{[m]}(x,\mu)\bar U(x,\mu). 
\end{eqnarray}
$V_{[m]}(x,\mu)$ is 
the instanton configuration chosen as the representative of $\mathfrak{U}[m]$, 
\begin{equation}
V_{[m]}(x,\mu)=\exp\left\{-\frac{2\pi i}{{L}^2}\left[{L}\delta_{\tilde
 x_{\mu},{L}-1}\sum_{\nu >\mu}m_{\mu\nu}\tilde
 x_{\nu}+\sum_{\nu<\mu}m_{\mu\nu}\tilde x_{\nu}\right]\right\},
\end{equation}
where $\tilde x_{\mu}= x_{\mu} \,{\rm  mod}\, L$. 
$\bar U(x,\mu)$ is to reproduce the field tensor
subtracted by the contribution from the instanton:
\begin{equation}
  \bar F_{\mu\nu}(x) = F_{\mu\nu}(x) - 2\pi \frac{m_{\mu\nu}}{L^2}.
\end{equation}
$\bar U(x,\mu)$ can be chosen in the complete axial gauge as 
\begin{eqnarray}
\bar U(x,\mu)=\prod_{\nu=\mu+1,...,n}
{\rm exp}\left(-{i\,\delta_{x_{\mu},L/2-1}
\sum_{y_{\nu}=0}^{x_{\nu}-1}
{}^\prime \sum_{y_{\mu}
={-L/2}}^{L/2-1}\bar F_{\mu\nu}(y_{\mu},y_{\nu})}\right) \ \ \ \ \nonumber\\
\times\prod_{\bar F \,\in\, S(0,x;\mu)} \,{\rm
 exp}({i \bar F_{\rho\sigma}(z)})\qquad \qquad
{\rm for}\,\,x \in \Gamma_n
\end{eqnarray}
where
$(y_{\mu},y_{\nu})=(0,...,0,y_{\mu},x_{\mu+1},...,
x_{\nu-1},y_{\nu},0,...,0)$
and $\sum^\prime$ is defined by
\begin{equation}
\sum_{t_i =0}^{x_i-1}{}^\prime f(x)
=\left\{ 
\begin{array}{ll}
\sum_{t_i=0}^{x_i-1} f(x) & (x_i \ge 1 )\\
0 & (x_i =0 )\\
\sum_{t_i=x_i}^{-1} (-1) f(x) & (x_i \le -1 )
\end{array}
 \right. .
\end{equation}
$U_{[w]}(x,\mu)$ is responsible for 
the $d$ non-contractible Wilson loops and defined by
\begin{eqnarray}
U_{[w]}(x,\mu)
=\left\{ 
\begin{array}{ll}
w_{\mu}\hspace{7mm} {\rm for}\,\,x_{\mu}=0\,{\rm mod} \,L,\\
0 \hspace{10mm} {\rm otherwise},\\
\end{array}
 \right. 
\end{eqnarray}
where $w_{\mu} \in U(1)$.

The components of the subtracted field tensor, 
$\bar F_{\mu\nu}(x)$, which appears in the above expression
are the linear independent components of the field tensor 
and can be used as the independent degrees of freedom of the
admissible gauge field in each magnetic sector.

\subsection{Local composite field of admissible gauge fields on
the finite lattice}
\label{subsect:lolcality in finite lattice}
In the finite lattice, the notion of the local composite field
of the admissible gauge field should be reconsidered. 
Here we adopt the definition of the local composite fields 
in the finite-volume lattice given in \cite{Kadoh:2003ii}.

\vspace{1em}
\noindent
{\it Local composite fields in the finite lattice}:\\
If $\phi(x)$ is a local composite field in the finite lattice,
there exist 
two local composite fields defined on the infinite lattice
$\phi_{\infty}(x)$ and $\Delta \phi_{\infty}(x)$ such that
\begin{eqnarray}
\phi(x)=\phi_{\infty}(x)+\Delta \phi_{\infty}(x), \ \ \ \ 
|\Delta \phi_{\infty}(x)|< c L^q e^{-L/2\varrho} ,
\end{eqnarray}
where the gauge fields in 
$\phi_{\infty}$ and $\Delta \phi_{\infty}$
are periodic. We also assume that $\phi(x)$ is expressed by 
the series expansion,
\begin{eqnarray}
\phi(x)=\sum_{k=1}^{\infty}\phi_k(x),
\label{eq:infinite_sum}
\end{eqnarray}
where $\phi_k(x)$ and their derivatives
$\phi_k(x;y_1,\nu_1;\cdots;y_m,\nu_m)$ with respect to 
the periodic gauge field
variables $U(y_1,\nu_1),\cdots,U(y_m,\nu_m)$ satisfy
\begin{eqnarray}
|\phi_k(x;y_1,\nu_1;\cdots;y_m,\nu_m)|
=\left\{ 
\begin{array}{ll}
\,0
&\qquad {\rm for} \,\,\, 2k < {\underset{z=y_1,\cdots,y_m}{\rm max}}
 \|x-z\|\\
\le  c_m k^{p_m}e^{-\theta k}
&\qquad {\rm otherwise}\\
\end{array}
 \right. 
\label{eq:exponentially-2}
\end{eqnarray}
with $m \ge 1$ and the constants 
$c_m,p_m >0$ and $\theta >0$ independent of the
gauge field. 

This locality property holds true for 
a gauge-invariant local composite field on the finite lattice
even when the field tensor is chosen as the independent variables.

\subsection{Modified lemma for the difference operators
on the finite lattice}
\label{subsec:lemma-finiteV}

In the finite-volume lattice, 
the difference operators are indtroduced in the same way
as in the infinite lattice given by eqs.~(\ref{eq:L}) 
and (\ref{eq:sum-L}), except 
that they are periodic with respect to $x,y$
and 
the summation ranges for $x,y$ are restricted to $\Gamma_n$.
Then the lemma corresponding to the 
Lemma~\ref{subsec:lemma-infiniteV} 
can be formulated by including finite-volume 
correction terms. The proof is given in appendix~\ref{app:lemma-finiteV}.

\vspace{1em} 
\noindent{\bf Lemma~\ref{subsec:lemma-finiteV}}\\
Let $L:\Omega_l \rightarrow \Omega_k \, (l\not=0)$ 
be a difference operator which satisfies
\begin{equation}
\hspace{-2cm}
{\rm d^\ast}L{\rm d}=0\,{\rm \,\,and\,\,}\,\sum
\hspace{-0.5mm}{\underset {1}{}}L {\rm d}=0
\,\,\,\,{\rm \,if\,\,\,}k=0 .
\end{equation}
Then there exist a form $\theta \in \Omega_{k+l}$ and three difference
operators $Q_0:\Omega_{l+1} \rightarrow \Omega_k$, $Q_1:\Omega_l \rightarrow 
\Omega_{k+1}$ and $\Delta L:\Omega_l \rightarrow \Omega_k$ such that
\begin{eqnarray}
L=\theta \circ +Q_1 {\rm d}+ {\rm d^\ast} Q_2 +\Delta L, 
\end{eqnarray}
where 
${\rm d^\ast}\theta=0$ and the coefficients of $\Delta L$ 
is a linear combination of those of $L$ with
$\|x-y\|\ge L/2$.\\

When $L$ is a gauge invariant operator, three operators $Q_1$, $Q_2$ and
$\Delta L$ and a field
$\theta$ are gauge invariant.
If the coefficients of $L$ are local with respect to $x$ or $y$ in terms
of subsect.\ref{subsect:lolcality in finite lattice} and satisfy the
following condition,
\begin{eqnarray}
|L_{\mu_1\cdots\mu_k\,,\,\nu_1\cdots\nu_l}(x,y)|< c(1+\|x-y\|^p){\rm
 e}^{-\|x-y\|/\varrho} \quad \ \ {\rm for}\,\,\,x-y \in \Gamma_n,
\end{eqnarray}
where $c,p$ and $\varrho$ are constants independent of the gauge
fields, then the coefficients of $Q_1$ and $Q_2$ are local with 
respect to $x$ or $y$ and satisfy 
\begin{eqnarray}
|Q_{i\,\,\mu_1\cdots\mu_k\,,\,\nu_1\cdots\nu_l}(x,y)|
< c_i(1+\|x-y\|^p){\rm
 e}^{-\|x-y\|/\varrho} \quad {\rm for}\,\,\,x-y \in \Gamma_n,
\, i=1,2, 
\end{eqnarray}
where $c_i$ are constants independent of the gauge fields, 
and $\theta$ is a local field.
The coefficients of $\Delta L$ satisfy
\begin{eqnarray}
|\Delta L_{\mu_1\cdots\mu_k\,,\,\nu_1\cdots\nu_l}(x,y)|< c^\prime 
L^p {\rm e}^{-L/2 \varrho} .
\end{eqnarray}
where $c^\prime$ is a constant independent of the gauge fields.

\subsection{Field tensor-based analysis of topological fields 
on the finite lattice}
\label{subsec:finite volume}

We first establish the following lemma through the field tensor-based method. 

\noindent{\bf Lemma \,\,\ref{subsec:finite volume}}\\
A gauge invariant local form $\phi \in \Omega_p$ which satisfies
\begin{eqnarray}
{\rm d^\ast} \phi(x)=0 \,{\rm \,\,and\,\,}\,\sum_{x \in \Gamma_n} 
\delta\phi(x)=0
\,\,\,\,{\rm \,if\,\,\,}p=0.
\end{eqnarray}
Then there exist three gauge invariant local fields, $\theta \in
\Omega_{p+2}$ which satisfies 
${\rm d^\ast}\theta=0$, $\omega \in \Omega_{p+1}$ 
and $\Delta \phi \in \Omega_{p}$, such that 
\begin{eqnarray}
\phi(x)=c_{[m,w]}(x)+\theta \circ \bar F(x) + {\rm d^\ast} \omega(x)
 +\Delta \phi(x),
\end{eqnarray}
where $c_{[m,w]}(x)$ is $\phi(x)$ 
for the gauge field $V_{[m]}(x,\mu)U_{[w]}(x,\mu)$. 
The coefficients of $\Delta \phi$ satisfy 
$|\Delta \phi_{\mu_1,...,\mu_k}(x)| \le \kappa 
{L}^{\sigma} {\rm e}^{-L/2\varrho}$, where
$\sigma$ and $\kappa$ are constants independent of the gauge field.

\vspace{1em}
\noindent {\sl Proof:}\ \ 
We rescale the independent components of the subtracted field tensor
$\bar F_{\mu\nu}(x) \in \mathfrak{\bar F}_{{}_x}$ by 
the continuous parameter $t \in [0,1]$, differentiate
$\phi(x)$ by $t$ and integrate it over the region $[0,1]$. 
Then we obtain
\begin{eqnarray}
\label{eq:difference-operator-L-V}
\phi(x)=c_{[m,w]}+ L \bar F(x),\ \ \ \ 
\end{eqnarray}
where $c_{[m,w]}=\phi(x)|_{\bar F=0}$ 
and $L$ is a gauge invariant operator which satisfies
the premise of the lemma \ref{subsec:lemma-finiteV}.
Then 
we obtain a solution to the constraint on $L$ 
with a finite-volume correction term.
\begin{eqnarray}
L=\theta \circ +Q_1 {\rm d}+ {\rm d^\ast} Q_2 +\Delta L,
\end{eqnarray}  
where 
$|\Delta L_{\mu_1\cdots\mu_k\,,\,\nu_1\cdots\nu_l}(x,y)|< \kappa^\prime 
L^{\sigma^\prime} {\rm e}^{-L/2 \varrho}$
with constants $\sigma^\prime$ and 
$\kappa^\prime$ independent of the gauge fields.
Then it follows
\begin{equation}
\phi=c_{[m,w]}+\theta \circ \bar F + {\rm d^\ast} \omega +\Delta \phi,
\end{equation}
where $\Delta \phi=\Delta L \bar F$. \quad  $\square$\\

Now by the several uses of the lemma \ref{subsec:finite volume},  
we can show that the topological field $q(x)$ is expressed as 
\begin{eqnarray}
\label{eq:final-result-finiteV}
&&\hspace{-5mm}q=
\sum_{k=0}^{[n/2]}
(\cdots((c^{(k)}_{[m,w]} \circ \bar F)\circ \bar F)\cdots \circ
\bar F)\circ \bar F + {\rm d^\ast}\bar\omega+\Delta  q , 
\end{eqnarray}
where $c^{(k)}_{[m,w]}(x)$ depends on the magnetic fluxes 
and the non-contractible Wilson loops
and $\vert
\Delta q \vert \le c {L}^{\sigma} {\rm e}^{-{L}/2\varrho}$. 
The topological properties of $q(x)$ implies 
$\sum_{x}\Delta q(x)=0$ and such a form is
always expressed as an exact form, 
$\Delta q={\rm d^\ast}\Delta \omega$. 
Then, without violating the locality property 
of the current $\bar\omega$ on the finite lattice, 
we can redefine $\bar\omega$ by including 
$\Delta\bar\omega$.
 
As the final step, we rewrite the above result
in terms of the orignal field tensor 
$F=\bar F+ \tilde m$
where $\tilde m=2\pi m_{\mu\nu}/{L^2}\,{\rm d}x_{\mu}{\rm d}x_{\nu}$. 
For this purpose, we note the following relation between
the coefficients $c^{(k)}_{[m,w]}(x)$ in eq.~(\ref{eq:final-result-finiteV})
and the coeffcients $c^{(k)}$ of the result in 
the infinite lattice, eq.~(\ref{eq:final-result-infiniteV}):
\begin{eqnarray}
\left|\,c^{(l)}_{[m,w]}(x)-\sum_{r=l}^{[n/2]}{}_r C {}_{r-l}\,
(\cdots((c^{(r)}\circ\tilde
 m )\circ \tilde m)\cdots \circ\tilde m)\, \right| \le \kappa_l
 L^{\sigma_l}e^{-L/2\varrho}.
\end{eqnarray}  
This relation is obtained by using the decomposition, 
$q(x)=q_\infty(x)+\Delta q_\infty(x)$,
and by comparing in detail the
algebraic construction of these coefficients.
The details of the proof of the four-dimensional case have 
been given in \cite{Kadoh:2003ii}.
Using this relation and $F=\bar F+ \tilde m$, we can rewrite 
$q(x)$ and obtain the final result
\begin{eqnarray}
&&\hspace{-5mm}q(x)=
\sum_{k=0}^{[n/2]}(\cdots((c^{(k)} \circ F)\circ F)\cdots \circ
 F)\circ F(x) + {\rm d^\ast}\omega(x) .
\end{eqnarray}

\section{Conclusion}

We have shown that 
the cohomological classification of the topological fields
in abelian lattice gauge theories can be achieved
by regarding the topological field as a gauge-invariant, local, function
of linear independent variables of the field tensor. This new method 
allows us to simplify the cohomological analysis by
reducing several steps 
both in the infinite lattice and the finite lattice.
This result will be usefull in the practical numerical 
implimentation of chiral lattice gauge theories.\cite{Kadoh:2004uu}
The application to other cases such as the local cohomology problem
in non-abelian lattice gauge theories is an open question. 
We leave this question for future study.

\acknowledgments

The authors would like to thank H.~Suzuki for valuable discussions.
They are also grateful to M.~L\"uscher for useful comments.
D.K. is supported in part by the Japan
Society for Promotion of Science under the Predoctoral
Research Program No.~15-887.


\appendix

\section{Poincar\'e lemma Lemmas in terms of differential forms :
  infinite lattice}
\label{app:lemma-infiniteV}

In the contractible space, any closed form is always exact. This fact is
well-known as the Poincar\'e lemma in the continuum theory and can be also
formulated  on the infinite lattice.\cite{Luscher:1998kn} 
The lemma can be also formulated for the general class of difference operators,
$L:\Omega_l
\rightarrow \Omega_{k}$, defined in the previous section, keeping locality
property of the operators.\cite{Luscher:1998kn,Kikukawa:2000kd} 
Here we simply quote the result. 

\vspace{1em}
\noindent{\bf Lemma~\ref{subsec:lemma-infiniteV}(a)
              \,\,\,(Poincar\'e lemma)} \\
Let $L$ be a difference operator $:\Omega_l \rightarrow \Omega_k$. 
Then 
\begin{eqnarray}
{\rm d}L=0\, \quad & \Rightarrow & \quad L=\delta_{kn}\sum\hspace{-0.5mm}{\underset {^1}{}}L + {\rm d} K,\label{eq:P1}\\
L{\rm d}=0\, \quad & \Rightarrow& \quad L=\delta_{l0}\sum\hspace{-0.5mm}{\underset {^2}{}}L +  K{\rm d},\label{eq:P2}\\
{\rm d^\ast}L=0 \quad & \Rightarrow& \quad L=\delta_{k0}\sum\hspace{-0.5mm}{\underset {^1}{}}L + {\rm d^\ast} K,\label{eq:P3}\\
L{\rm d^\ast}=0 \quad & \Rightarrow& \quad
 L=\delta_{ln}\sum\hspace{-0.5mm}{\underset {^2}{}}L 
+ K{\rm d^\ast},\label{eq:P4}
\end{eqnarray}
where $K$ is difference operator constructed from $L$. 
An important point to note is that $\sum_{1,2}L$ and $K$ are gauge invariant if
$L$ is gauge invariant. If a coefficient of the operator $L$ is local with
respect to $x$ or $y$, then one of the solution $K$ is local with
respect to $x$ or $y$.    
Moreover, instead of the compact support if the coefficient satisfies  
\begin{eqnarray}
|L_{\mu_1\cdots\mu_k\,,\,\nu_1\cdots\nu_l}(x,y)|< c_1(1+\|x-y\|^p){\rm e}^{-\|x-y\|/\varrho},
\end{eqnarray}
where $c_1,p$ and $\varrho$ are constants independent of the gauge
fields, then the coefficient of $K$ satisfies  
\begin{eqnarray}
|K_{\mu_1\cdots\mu_k\,,\,\nu_1\cdots\nu_l}(x,y)|< c_2(1+\|x-y\|^p){\rm e}^{-\|x-y\|/\varrho}.
\end{eqnarray}
where $c_2$ is a constant independent of the gauge fields.

The Poincar\'e lemma concerns the case of 
${\rm d^\ast} L=0$ or $L {\rm d}=0$. However, 
in the cohomological analysis of the topological field,
we often encounter 
a difference operator $L$ which satisfies ${\rm d^\ast} L{\rm d}=0$.
Then it turns out to be useful to formulate a lemma which 
tells us how to express such a difference operator.\cite{Fujiwara:1999fj}  
Here we give a proof of the lemma using the differential forms and
the $\circ$ product. 


\vspace{1em}
\noindent{\bf Lemma~\ref{subsec:lemma-infiniteV}(b)}\\
Let $L:\Omega_l (l \ne 0)\rightarrow \Omega_k$ be a difference operator which
satisfies\\
\begin{equation}
\hspace{-2cm}{\rm d^\ast}L{\rm d}=0
\quad {\text{and}} \quad \sum\hspace{-0.5mm}{\underset {1}{}} L {\rm d}=0
\,\,\,\,{\rm \,if\,\,\,}k=0
\end{equation}
Then there exist a form $\theta \in \Omega_{k+l}$ that satisfies ${\rm
d^\ast}\theta=0$ and two difference
operators, $Q_0:\Omega_{l+1} \rightarrow \Omega_k$ and $Q_1:\Omega_l \rightarrow 
\Omega_{k+1}$ such that
\begin{eqnarray}
L=\theta \circ +Q_0 {\rm d}+ {\rm d^\ast} Q_1.
\end{eqnarray}

If the coefficient of $L$ is local with respect $x$ or $y$ and satisfies
\begin{eqnarray}
|L_{\mu_1\cdots\mu_k\,,\,\nu_1\cdots\nu_l}(x,y)|< c(1+\|x-y\|^p){\rm e}^{-\|x-y\|/\varrho},
\end{eqnarray}
where $c,p$ and $\varrho$ are constants independent of the gauge
fields, then the coefficients of $Q_1$ and $Q_2$ are also local with 
respect $x$ or $y$ and satisfy  
\begin{eqnarray}
|Q_{i\,\,\mu_1\cdots\mu_k\,,\,\nu_1\cdots\nu_l}(x,y)|< c_i(1+\|x-y\|^p){\rm e}^{-\|x-y\|/\varrho},
\end{eqnarray}
where $c_i(i=1,2)$ are constants independent of the gauge fields, and
$\theta $ is a local field.

\noindent {\sl Proof:} \ \
Here we consider the case
of $k+l \le n$, but it is easy to extend the case of $k+l > n$
Applying the Poincar\'e lemma repeatedly, 
the following chain is obtained,
\begin{eqnarray}
L^{(0)}\,{\rm d} &=&{\rm
 d^\ast}L^{(1)},\hspace{5cm}\label{eq:cahin-1}\\
L^{(1)}\,{\rm d}&=&{\rm d^\ast}L^{(2)},\\
:\nonumber\\
:\nonumber\\
L^{(l-2)}\,{\rm d}&=&{\rm d^\ast}L^{(l-1)},\\
L^{(l-1)}\,{\rm d}&=&{\rm d^\ast}L^{(l)},
\label{eq:condition1}\\
L^{(l)}\,{\rm d}&=&0,
\end{eqnarray}
where $L = L^{(0)}$ and $L^{(m)}:\Omega_{l-m} \rightarrow \Omega_{k+m}$.  
Then the Poincar\'e lemma allow us to conclude that there exists a difference
operator $Q^{(l)}:\Omega_{1} \rightarrow \Omega_{k+l}$ such that
\begin{equation}
L^{(l)}=\sum\hspace{-0.5mm}{\underset {^2}{}}L + B^{(l)}\,{\rm d},
\label{eq:poincare}
\end{equation} 
and we may express $\sum\hspace{-0.5mm}{\underset {^2}{}}L = \theta \circ $, \,$\theta \in \Omega_{k+l}$.
By acting eq.(\ref{eq:condition1}) on a constant, we can assert that
${\rm d^\ast} \theta =0$. Then using 
eq.(\ref{eq:circle-prod-property}), eq.(\ref{eq:condition1}) is
rewritten as follows:
\begin{equation}
(L^{(l-1)}- \theta \circ-{\rm d^\ast} Q^{(l)})\,{\rm d}=0.
\label{eq:condition-closed}
\end{equation}
where an extra sign in the second equation of
 eq.(\ref{eq:circle-prod-property}) is absorbed into 
a redefinition of $\theta$.
Again the Poincar\'e lemma can be used to obtain
\begin{equation}
L^{(l-1)}=\theta \circ+{\rm d^\ast} Q^{(l)}+Q^{(l-1)}{\rm d}.
\end{equation} 
where $Q^{(l-1)}:\Omega_{2} \rightarrow \Omega_{k+l-1}$.
Repeating the same process, we finally obtain
\begin{equation}
L^{(0)}=\theta \circ+{\rm d^\ast} Q^{(1)}+Q^{(0)}{\rm d},
\end{equation}
where ${\rm d^\ast} \theta =0$, $Q^{(1)}:\Omega_{l} \rightarrow \Omega_{k+1} $ and $Q^{(0)}:\Omega_{l+1} \rightarrow \Omega_k$.
For $k+l > n$ we have $\theta=0$. It is obvious that the statement with
respect to the locality satisfies because the solutions of the Poincar\'e
lemma always have same locality
properties.
\ \ \ \ $\square$

\section{Modified Poincar\'e lemmas in terms of differential forms :
finite lattice} 
\label{app:lemma-finiteV}

On n-dimensional torus in the continuum,  a closed form
can not be always expressed as an exact form globally because the cohomology 
group of the space is non-trivial. Then, if one consider the finite periodic lattice
with the linear extent $L$, 
the Poincar\'e lemma formulated in the previous section does not hold anymore
in general.
However, we can show that the lattice counterpart of 
the corollary of de Rham theorem holds true and moreover, 
for a sufficiently large (finite) lattice, a modified version of 
the lemma holds with a small correction term suppressed 
exponentially in lattice size. 
These lemmas can substitute the Poincar\'e lemma in 
the infinite lattice and allows us to perform the cohomological analysis
directly on the finite lattice.\cite{Kadoh:2003ii}
Here we simply quote those results. 

\vspace{1em}
\noindent{\bf Lemma~\ref{subsec:lemma-finiteV}(a)\,\,(Modified
Poincar\'e lemma)} \\
Let $L:\Omega_l \rightarrow \Omega_k$ be a difference operator which
satisfies
\begin{eqnarray}
{\rm d}L=0 \,\quad &s.t.& \quad L=\delta_{kn}\sum\hspace{-0.5mm}{\underset {^1}{}} L + {\rm d} K+\Delta L,\label{eq:MP1}\\
L{\rm d}=0 \,\quad &s.t.& \quad L=\delta_{l0}\sum\hspace{-0.5mm}{\underset {^2}{}} L_+  K{\rm d}+\Delta L,\label{eq:MP2}\\
{\rm d^\ast}L=0 \quad &s.t.&\quad L=\delta_{k0}\sum\hspace{-0.5mm}{\underset {^1}{}} L + {\rm d^\ast} K+\Delta L,\label{eq:MP3}\\
L{\rm d^\ast}=0 \quad &s.t.&\quad L=\delta_{ln}\sum\hspace{-0.5mm}{\underset {^2}{}} L + K{\rm d^\ast}+\Delta L\label{eq:MP4},
\end{eqnarray}
where $K,\sum_{1,2}L$ and $\Delta L$ are difference operators. 
Moreover $\Delta L$ is constructed from a linear summation of $L$ where
$\|x-y\| \ge L/2 $.
$\Delta L=0$ when the terms $\sum_{1,2}L$ exist.\\

\vspace{1em}
\noindent{\bf Lemma~\ref{subsec:lemma-finiteV}(b)
 \,\,(Corollary of de Rham theorem)}\\
Let $L:\Omega_l \rightarrow \Omega_k$ be a difference operator which
 satisfies
\begin{eqnarray}
{\rm d}L=0,\,\,\,\sum\hspace{-0.5mm}{\underset {^1}{}} L=0 \,\quad& s.t.&\quad L={\rm d} K,\\
L{\rm d}=0,\,\,\,\sum\hspace{-0.5mm}{\underset {^2}{}}L=0 \,\quad & s.t.&\quad L= K{\rm d},\\
{\rm d^\ast}L=0,\,\,\,\sum\hspace{-0.5mm}{\underset {^1}{}}L=0 \quad & s.t.
&\quad L={\rm d^\ast} K,\label{eq:de-Rham-3}\\
L{\rm d^\ast}=0,\,\,\,\sum\hspace{-0.5mm}{\underset {^2}{}} L=0 \quad & s.t.&\quad L=K{\rm d^\ast},
\end{eqnarray}
where $K$ is a difference operator constructed from $L$. \\

For lemma \ref{subsec:lemma-finiteV} (a) and (b),
if the coefficient of $L$ is local with respect to $x$ or $y$ and satisfies
\begin{eqnarray}
|L_{\mu_1\cdots\mu_k\,,\,\nu_1\cdots\nu_l}(x,y)|< c_1(1+\|x-y\|^p){\rm e}^{-\|x-y\|/\varrho},
\end{eqnarray}
where $c_1,p$ and $\varrho$ are constants independent of the gauge
fields, then the coefficients of $K$ is also local with 
respect to $x$ or $y$ and satisfies
\begin{eqnarray}
&& | K_{\mu_1\cdots\mu_k\,,\,\nu_1\cdots\nu_l}(x,y)|< c_2(1+\|x-y\|^p){\rm
 e}^{-\|x-y\|/\varrho},
\end{eqnarray}
and the coefficients of $\Delta L$ satisfies
\begin{eqnarray}
&& |\Delta L_{\mu_1\cdots\mu_k\,,\,\nu_1\cdots\nu_l}(x,y)|
< c_3{L^p}{\rm e}^{-L/2\varrho}
\end{eqnarray}
where $c_2$ and $c_3$ and constants independent of the gauge fields.\\

A lemma which corresponds to 
Lemma~\ref{subsec:lemma-infiniteV}(b) can be formulated 
in the infinite lattice. 

\vspace{1em} 
\noindent{\bf Lemma~\ref{subsec:lemma-finiteV}(c)}\\
Let $L:\Omega_l \rightarrow \Omega_k$ be a difference operator which
satisfies
\begin{equation}
\hspace{-2cm}{\rm d^\ast}L{\rm d}=0\,{\rm \,\,and\,\,}\,\sum L {\rm d}=0
\,\,\,\,{\rm \,if\,\,\,}k=0 .
\end{equation}
Then there exist a form $\theta \in \Omega_{k+l}$ and three difference
operators $Q_0:\Omega_{l+1} \rightarrow \Omega_k$, $Q_1:\Omega_l \rightarrow 
\Omega_{k+1}$ and $\Delta L:\Omega_l \rightarrow \Omega_k$ such that
\begin{eqnarray}
L=\theta \circ +Q_1 {\rm d}+ {\rm d^\ast} Q_2 +\Delta L, 
\end{eqnarray}
where 
${\rm d^\ast}\theta=0$ and $\Delta L$ is 
constructed from a linear summation of $L$ where $\|x-y\| \ge L/2 $.\\

For lemma \ref{subsec:lemma-finiteV}(c), if $L$ satisfies the conditions which are mentioned below lemma
\ref{subsec:lemma-finiteV}(b), then $K$ and $\Delta L$ also satisfy the
conditions mentioned there and $\theta$ is a local field.\\

\noindent {\sl Proof:} \ \ 
The strategy of the proof is almost same as the 
case of lemma~\ref{subsec:lemma-infiniteV}(b). 
However the existence of the
finite size correction leads us to extra difficulties. If the difference 
operator $L^{(m)}$ satisfies the lemma's assumption, 
${\rm d^\ast}L^{(m)}{\rm d}=0$, applying the modified Poincar\'e 
lemma eq.(\ref{eq:MP3})  we conclude that
\begin{eqnarray}
L^{(m)}{\rm d} =
{\rm d}^\ast \bar L^{(m+1)} +\Delta \bar L^{(m+1)}.\label{eq:finite-1}
\end{eqnarray}
Now we have to redefine the operator $\bar L^{(m+1)} $ appropriately 
in order to use the lemma's assumption repeatedly. Anyway,
multiplying ${\rm d}$ to eq.(\ref{eq:finite-1}) from right we obtain
\begin{equation}
{\rm d}^\ast \bar L^{(m+1)} {\rm d}=-\Delta \bar L^{(m+1)}{\rm d}. 
\label{eq:finite-2}
\end{equation} 
It is obvious that the operator $\Delta \bar L^{(m+1)}{\rm d}$ satisfies 
the premise of the corollary of de Rham theorem. We can change the right 
hand side of eq.(\ref{eq:finite-2}) in the form as 
${\rm d}^\ast \Delta K {\rm d}$, however, a simple application of the 
lemma (\ref{eq:de-Rham-3}) seems to lead that ${\rm d}$ vanishes instead 
of appearing $d^\ast$. It relates how to construct a solution of the 
Poincar\'e lemma. In general, when we construct a typical solution of 
the Poincare lemma, for example, for the case of ${\rm d}^\ast
L(x,y)=0$, 
the solution $K(x,y)$ is obtained by integrating $L(x,y)$ over $x$. 
To preserve the locality, $y$ should be chosen as the reference point 
that corresponds an initial point of the integral. It causes the fact 
that the equation ${\rm d}^\ast (L{\rm d})=0$ can not be solved as the 
form $L {\rm d}={\rm d}^\ast K{\rm d}$ without violating the locality 
property. However since the operator $\Delta \bar L^{(m+1)}{\rm d}$
is exponentially small, it is not necessary to take care of the 
locality property and 
the reference point can be taken arbitrary. Therefore the difference 
operator $\Delta L^{(m+1)} {\rm d}$ can be  expressed as the following form,
\begin{equation}
\Delta \bar L^{(m+1)} {\rm d}={\rm d}^\ast \Delta K^{(m+1)} {\rm d}.
\end{equation}
Then we introduce new difference operators $L^{(m+1)} $ and 
$\Delta R^{(m+1)} $ as
\begin{eqnarray}
L^{(m+1)} =\bar L^{(m+1)}+\Delta K^{(m+1)}, \ \ \ \ 
\Delta R^{(m+1)} =\Delta \bar L^{(m+1)}-{\rm d}^\ast \Delta K^{(m+1)}.
\end{eqnarray}
Thus, instead of eq.(\ref{eq:finite-1}), we obtain the following equations
\begin{eqnarray}
L^{(m)}{\rm d} ={\rm d}^\ast L^{(m+1)} +\Delta R^{(m+1)},
 \ \ \ \ {\rm d}^\ast  L^{(m+1)}{\rm d}=0.
\end{eqnarray}
We can write down the chain for the case of $k+l \le n$.
\begin{eqnarray}
L^{(0)}\,{\rm d} &=&{\rm
 d^\ast}L^{(1)}+\Delta R^{(1)}, \ \ \ \ \ \ \ \  {\rm d}^\ast  L^{(1)}{\rm d}=0,\label{eq:chain2-1}\hspace{5cm}\\
L^{(1)}\,{\rm d}&=&{\rm d^\ast}L^{(2)}+\Delta R^{(2)}, \ \ \ \ \ \ \ \ {\rm d}^\ast  L^{(2)}{\rm d}=0,\\
:\nonumber\\
:\nonumber\\
L^{(l-2)}\,{\rm d}&=&{\rm d^\ast}L^{(l-1)}+\Delta R^{(l-1)}, \ \ \ \ {\rm d}^\ast  L^{(l-1)}{\rm d}=0,\\
L^{(l-1)}\,{\rm d}&=&{\rm d^\ast}L^{(l)}+\Delta R^{(l)}, 
\ \ \ \  \ \ \ \ \ \ \sum\hspace{-0.5mm}{\underset {^2}{}} {\rm d}^\ast  L^{(l)}=0,
\label{eq:finite-condition1}\\
L^{(l)}\,{\rm d}&=&0.
\end{eqnarray}
Applying the modified Poincar\'e lemma eq.(\ref{eq:MP2}) to the last line of the chain, we get the same equation to the case of the infinite volume lattice,
\begin{equation}
L^{(l)}=\theta \circ +Q^{(l)}{\rm d}
\end{equation}
where $\theta \in \Omega_{k+l}$. For $\Delta R^{(l)}$, we apply the corollary of de Rham theorem and  get 
\begin{equation}
\Delta R^{(l)}=\Delta B^{(l)}{\rm d}.  
\end{equation} 
Acting eq.(\ref{eq:finite-condition1}) on a constant, $\theta$ satisfies ${\rm d}^\ast \theta=0$. Therefore eq.(\ref{eq:finite-condition1}) can be expressed in the form,
\begin{equation}
(L^{(l-1)}- \theta \circ-{\rm d^\ast} Q^{(l)}-\Delta B^{(l)})\,{\rm d}=0.
\end{equation}
where we redefine $(-1)^{k+l} \theta$ to $\theta$. Again we apply the modified Poincar\'e lemma eq.(\ref{eq:MP2}) and redefine $\Delta B^{(l)}$ and $\Delta L^{(l-1)}$, which is an exponentially small correction that arises from eq.(\ref{eq:MP2}), to $\Delta L^{(l-1)}$. Then we obtain
\begin{equation} 
L^{(l-1)}= \theta \circ+{\rm d^\ast} Q^{(l)}+Q^{(l-1)}{\rm d}+\Delta L^{(l-1)}.
\end{equation}
Thus we carry out above procedure repeatedly and finally obtain
\begin{equation}
L^{(0)}= \theta \circ+{\rm d^\ast} Q^{(1)}+Q^{(0)}{\rm d}+\Delta L^{(0)}.
\end{equation}
where ${\rm d}^\ast \theta=0$. For the case of $k+l>n$, the chain stops at $n-k-1$ line and we have vanishing $\theta$.  \quad  $\square$

\end{document}